\documentstyle[prl,floats,twocolumn,epsf,psfrag,aps]{revtex}

\tighten
\begin{document}
\draft
\twocolumn[\hsize\textwidth\columnwidth\hsize\csname @twocolumnfalse\endcsname

\title{Unconditional teleportation of continuous-variable entanglement}
\author{P.\ van Loock$^1$ and Samuel L.\ Braunstein$^{1,2}$}
\address{$^1$SEECS, University of Wales, Bangor LL57 1UT, UK\\
$^2$Hewlett-Packard Labs, Mail Box M48, Bristol BS34 8QZ, UK}
\maketitle

\begin{abstract} 
We give a protocol and criteria for demonstrating unconditional teleportation
of continuous-variable entanglement (i.e., entanglement swapping).
The initial entangled states are produced with squeezed
light and linear optics.
We show that any nonzero entanglement (any nonzero squeezing)
in both of two entanglement sources is sufficient for entanglement
swapping to occur. In fact, realization of continuous-variable 
entanglement swapping is possible using only {\it two} single-mode 
squeezed states.
\end{abstract}
\vspace{3ex}
]

Quantum teleportation enables reliable transportation of
quantum information encoded in non-orthogonal quantum states.
It is only possible with entanglement.
Teleportation was originally proposed for discrete variables \cite{Benn} 
and later also for continuous variables \cite{Vaid,Sam98a}.
Discrete-variable teleportation has been performed experimentally
for single-photon polarization states \cite{Bou,Mart}. 
Continuous-variable teleportation has been realized for
coherent states of electromagnetic field modes \cite{Furu}. 
But coherent states, although non-orthogonal, are very close to classical 
states. A real challenge for quantum teleportation is the
teleportation of truly non-classical states like entangled states.
This `entanglement swapping' was first introduced in the context 
of single-photon polarization states \cite{Zuk}.
It means to entangle two quantum systems
that have never directly interacted with each other. 
With single photons, it has already been demonstrated experimentally 
\cite{Pan}. Practical uses of entanglement swapping have been suggested
\cite{Bose1,Bose2,Briegel,Duer} and it has also been generalized for 
multiparticle systems \cite{Bose1}. All these investigations 
have only referred to discrete-variable systems, namely two-level systems. 
We will demonstrate that entanglement swapping can also be realized in 
continuous-variable systems where the source of entanglement is 
two-mode squeezed light. In contrast to the scheme of 
Polkinghorne and Ralph \cite{Ralph} where polarization-entangled states 
of single photons are teleported using squeezed-state entanglement, 
in our scheme both entangled states are produced with squeezed 
light. This enables {\it unconditional} teleportation of entanglement
{\it without} post-selection of `successful' events by photon detections.
Unconditional teleportation of continuous-variable entanglement
has been independently investigated by Tan \cite{Tan}.
We will compare Tan's results with ours at the end.

Due to the finite degree of entanglement arising from squeezed states,
the entanglement that emerges from entanglement swapping
is never as good as the entanglement of the two initial entanglement 
sources. However, entanglement swapping as here proposed occurs every 
inverse bandwidth time and is very efficient (near unit efficiency).
The fidelity criterion for coherent-state teleportation \cite{Fuchs}
will enable us to recognize the entanglement produced from
entanglement swapping. The maximum average fidelity achievable
using the output of entanglement swapping for teleportation 
indicates the quality of the entanglement from entanglement swapping.
Applying this operational quality criterion for entanglement gives us 
also a protocol for the experimental verification of 
entanglement swapping.      

For our entanglement swapping scheme
(Fig.~1), we need two entangled states of the electromagnetic field:
a two-mode squeezed state of mode 1 and mode 2 and a two-mode squeezed state 
of mode 3 and mode 4. This can be described in the Heisenberg representation 
by
\begin{eqnarray}\label{HeisEPR}
\hat{x}_1&=&(e^{+r_1} \hat{x}^{(0)}_1+ e^{-r_2} \hat{x}^{(0)}_2)/\sqrt{2} \;,
\nonumber\\
\hat{p}_1&=&(e^{-r_1} \hat{p}^{(0)}_1+ e^{+r_2} \hat{p}^{(0)}_2)/\sqrt{2} \;,
\nonumber\\
\hat{x}_2&=&(e^{+r_1} \hat{x}^{(0)}_1- e^{-r_2} \hat{x}^{(0)}_2)/\sqrt{2} \;,
\nonumber\\
\hat{p}_2&=&(e^{-r_1} \hat{p}^{(0)}_1- e^{+r_2} \hat{p}^{(0)}_2)/\sqrt{2} \;,
\nonumber\\
\hat{x}_3&=&(e^{+s_1} \hat{x}^{(0)}_3+ e^{-s_2} \hat{x}^{(0)}_4)/\sqrt{2} \;,
\nonumber\\
\hat{p}_3&=&(e^{-s_1} \hat{p}^{(0)}_3+ e^{+s_2} \hat{p}^{(0)}_4)/\sqrt{2} \;,
\nonumber\\
\hat{x}_4&=&(e^{+s_1} \hat{x}^{(0)}_3- e^{-s_2} \hat{x}^{(0)}_4)/\sqrt{2} \;,
\nonumber\\
\hat{p}_4&=&(e^{-s_1} \hat{p}^{(0)}_3- e^{+s_2} \hat{p}^{(0)}_4)/\sqrt{2} \;,
\end{eqnarray}
where a superscript `$(0)$' denotes initial vacuum modes.
The operators $\hat{x}$ and $\hat{p}$ represent the electric quadrature 
amplitudes (the real and imaginary part of the mode's annihilation operator).
These two-mode squeezed vacuum states can be generated either 
directly as the output of a nondegenerate optical parametric amplifier 
\cite{Reid} or by combining two squeezed vacuum modes at a beamsplitter
(see Fig.~1).
But note that the two independently squeezed single-mode 
states combined at a beamsplitter to create entanglement need not be
equally squeezed. In fact, even one single-mode squeezed state
superimposed with vacuum yields an entangled two-mode state 
\cite{Ah,Matteo} enabling quantum teleportation \cite{Pvl}. Thus, we use 
four different squeezing parameters $r_1$, $r_2$, $s_1$ and $s_2$.

\begin{figure}[htb]
\begin{center}
\begin{psfrags}
     \psfrag{Alice}{\Large \bf Alice}
     \psfrag{Bob}{\Large \bf Bob}
     \psfrag{Claire}{\Large \bf Claire}
     \psfrag{1}{\Large 1}
     \psfrag{2}{\Large 2}
     \psfrag{3}{\Large 3}
     \psfrag{4}{\Large 4}
     \psfrag{x}{\Large ${\rm D}_x~~~~~$}
     \psfrag{p}{\Large ${\rm D}_p~~~~~$}
     \psfrag{u}{\Large u}
     \psfrag{v}{\Large v$~~$} 
\epsfxsize=2.4in
     \epsfbox[-60 60 400 400]{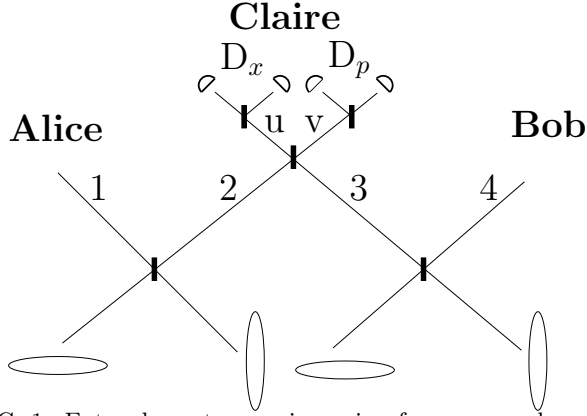}
\end{psfrags}
\end{center}
\caption{Entanglement swapping using four squeezed vacuum states.
         Before the detections, mode 1 is entangled with mode 2 and
         mode 3 is entangled with mode 4.} 
\label{fig1}
\end{figure}

Let us introduce ``Alice'', ``Bob'' and ``Claire'' to illustrate
the whole protocol with entanglement swapping and subsequent teleportation
of a coherent state. Alice and Claire shall share the entangled state of 
mode 1 and 2 while Claire and Bob are sharing the other entangled state 
of mode 3 and 4 (Fig.~1). Thus, initially Alice and Bob do not share an 
entangled state. Now Alice wants to teleport an unknown 
coherent state to Bob and asks Claire for her assistance.
Claire combines mode 2 and mode 3 at a beamsplitter and detects the
quadratures 
$\hat{x}_{\rm u}=(\hat{x}_2-\hat{x}_3)/\sqrt{2}$,
$\hat{p}_{\rm v}=(\hat{p}_2+\hat{p}_3)/\sqrt{2}$.
Let us write Bob's mode 4 as
\begin{eqnarray}\label{mode4}
\hat{x}_4&=&\hat{x}_2-(\hat{x}_3-\hat{x}_4)
-\sqrt{2}\hat{x}_{\rm u} \;,\nonumber\\
\hat{p}_4&=&\hat{p}_2+(\hat{p}_3+\hat{p}_4)
-\sqrt{2}\hat{p}_{\rm v} \;,
\end{eqnarray}
and Alice's mode 1 as
\begin{eqnarray}\label{mode1}
\hat{x}_1&=&\hat{x}_3+(\hat{x}_1-\hat{x}_2)
+\sqrt{2}\hat{x}_{\rm u} \;,\nonumber\\
\hat{p}_1&=&\hat{p}_3+(\hat{p}_1+\hat{p}_2)
-\sqrt{2}\hat{p}_{\rm v} \;.
\end{eqnarray}
Claire's detection yields classical results $x_{\rm u}$ and 
$p_{\rm v}$. Bob's mode 4 in Eqs.~(\ref{mode4}) and Alice's mode 1
in Eqs.~(\ref{mode1}) then become
\begin{eqnarray}\label{1and4}
\hat{x}_4&=&\hat{x}_2-\sqrt{2}e^{-s_2} \hat{x}^{(0)}_4
-\sqrt{2}x_{\rm u} \;,
\nonumber\\
\hat{p}_4&=&\hat{p}_2+\sqrt{2}e^{-s_1} \hat{p}^{(0)}_3
-\sqrt{2}p_{\rm v} \;,
\nonumber\\
\hat{x}_1&=&\hat{x}_3+\sqrt{2}e^{-r_2} \hat{x}^{(0)}_2
+\sqrt{2}x_{\rm u} \;,
\nonumber\\
\hat{p}_1&=&\hat{p}_3+\sqrt{2}e^{-r_1} \hat{p}^{(0)}_1
-\sqrt{2}p_{\rm v} \;.
\end{eqnarray} 
For $s_1=s_2=s\to\infty$, the quadrature operators of mode 4 become
those of mode 2 up to a (random) classical phase-space displacement.
In every single projection, mode 4 gets entangled with mode 1 as 
mode 2 has been before.
For $r_1=r_2=r\to\infty$, the quadrature operators of mode 1 become
those of mode 3 up to a (random) classical phase-space displacement.
In every single projection, mode 1 gets entangled with mode 4 as mode 3 
has been before.
Mode 2 is perfectly teleported to mode 4 ($s\to\infty$) or mode 3 is 
perfectly teleported to mode 1 ($r\to\infty$) apart from local
classical displacements. 
The entanglement of one of the initial two-mode squeezed states
is completely preserved for infinite squeezing in the other two-mode
squeezed state. 
But also for any nonzero squeezing in both initial entanglement
sources, Claire's detection of mode 2 and 3 projects mode 1 and 4 
on inseparable states \cite{Peter}. 
However, we will see that Alice and Bob cannot use
mode 1 and 4 for subsequent quantum teleportation 
without information about Claire's measurement
results. At least one of them, Alice or Bob, has to receive
from Claire the information that the detection of mode 2 and 3
has been performed and its results.
Finally, the entanglement of the single projections is `unwrapped'
via appropriate displacements.
Let us assume Bob obtains the classical results from Claire. 
Now Bob can displace mode 4 as 
$\hat{x}_4\longrightarrow\hat{x}_4'=\hat{x}_4+g_{\rm swap}
\sqrt{2} x_{\rm u}$,
$\hat{p}_4\longrightarrow\hat{p}_4'=\hat{p}_4+g_{\rm swap}
\sqrt{2} p_{\rm v}$.
The parameter $g_{\rm swap}$ describes a normalized gain.
Bob's mode then becomes
\begin{eqnarray}\label{mode4'}
\hat{x}_4'&=&\frac{g_{\rm swap}}{\sqrt{2}}e^{+r_1} \hat{x}^{(0)}_1
-\frac{g_{\rm swap}}{\sqrt{2}}e^{-r_2} \hat{x}^{(0)}_2
\nonumber\\
&-&\frac{g_{\rm swap}-1}{\sqrt{2}}
e^{+s_1}\hat{x}^{(0)}_3-\frac{g_{\rm swap}+1}{\sqrt{2}}
e^{-s_2}\hat{x}^{(0)}_4 \;,
\nonumber\\
\hat{p}_4'&=&\frac{g_{\rm swap}}{\sqrt{2}}e^{-r_1} \hat{p}^{(0)}_1 
-\frac{g_{\rm swap}}{\sqrt{2}}e^{+r_2} \hat{p}^{(0)}_2
\nonumber\\
&+&\frac{g_{\rm swap}+1}{\sqrt{2}} e^{-s_1}
\hat{p}^{(0)}_3+\frac{g_{\rm swap}-1}{\sqrt{2}}
e^{+s_2}\hat{p}^{(0)}_4 \;.
\end{eqnarray}
As in `usual' teleportation, Alice now couples the unknown input state
she wants to teleport to Bob (described by $\hat{x}_{\rm in}$, 
$\hat{p}_{\rm in}$) with her mode 1 at a beamsplitter
and measures the superpositions
$\hat{x}_{\rm u}'=(\hat{x}_{\rm in}-\hat{x}_1)/\sqrt{2}$,
$\hat{p}_{\rm v}'=(\hat{p}_{\rm in}+\hat{p}_1)/\sqrt{2}$.
Based on the classical results sent to him from Alice, Bob displaces
his `new' mode $4'$,
$\hat{x}_4'\longrightarrow\hat{x}_{\rm tel}=\hat{x}_4'+g\sqrt{2}
x_{\rm u}'$,
$\hat{p}_4'\longrightarrow\hat{p}_{\rm tel}=\hat{p}_4'+g\sqrt{2}
p_{\rm v}'$,
with the gain $g$. For $g=1$ and nonunit detector efficiencies, 
Bob's outgoing mode then becomes
\begin{eqnarray}\label{tel}
\hat{x}_{\rm tel}=&&\hat{x}_{\rm in}
+\frac{g_{\rm swap}-1}{\sqrt{2}}e^{+r_1} \hat{x}^{(0)}_1
-\frac{g_{\rm swap}+1}{\sqrt{2}}e^{-r_2} \hat{x}^{(0)}_2
\\
-&&\frac{g_{\rm swap}-1}{\sqrt{2}}
e^{+s_1}\hat{x}^{(0)}_3-\frac{g_{\rm swap}+1}{\sqrt{2}}
e^{-s_2}\hat{x}^{(0)}_4
\nonumber\\
+g_{\rm swap}&&\sqrt{\eta_c^{-2}-1}\;
(\hat{x}^{(0)}_d+\hat{x}^{(0)}_e)
+\sqrt{\eta_a^{-2}-1}\;(\hat{x}^{(0)}_f+\hat{x}^{(0)}_g),
\nonumber\\
\hat{p}_{\rm tel}=&&\hat{p}_{\rm in}
+\frac{g_{\rm swap}+1}{\sqrt{2}}e^{-r_1} \hat{p}^{(0)}_1 
-\frac{g_{\rm swap}-1}{\sqrt{2}}e^{+r_2} \hat{p}^{(0)}_2
\nonumber\\
+&&\frac{g_{\rm swap}+1}{\sqrt{2}}
e^{-s_1}\hat{p}^{(0)}_3+\frac{g_{\rm swap}-1}{\sqrt{2}}
e^{+s_2}\hat{p}^{(0)}_4
\nonumber\\
+g_{\rm swap}&&\sqrt{\eta_c^{-2}-1}\;
(\hat{p}^{(0)}_k+\hat{p}^{(0)}_l)
+\sqrt{\eta_a^{-2}-1}\;(\hat{p}^{(0)}_m+\hat{p}^{(0)}_n).\nonumber
\end{eqnarray}
The parameters $\eta_c$ and $\eta_a$ describe detector efficiencies
in Claire's and Alice's detections, respectively.
Note that for $g_{\rm swap}=1$, Bob's teleported mode from 
Eqs.~(\ref{tel}) is the same as if Alice teleports
her input state to Claire with unit gain using the entangled state of
mode 1 and 2, and Claire teleports the resulting output state 
to Bob with unit gain using the entangled state of mode 3 and 4. 

The teleportation fidelity for a coherent state 
input $\alpha_{\rm in}$, defined as 
$F\equiv\langle\alpha_{\rm in}|\hat{\rho}_{\rm tel}|
\alpha_{\rm in}\rangle=\pi Q_{\rm tel}(\alpha_{\rm in})$ \cite{Fuchs},
is
\begin{eqnarray}\label{fid1}
F=\frac{1}{2\sqrt{\sigma_x\sigma_p}}\exp\left
[-(1-g)^2\left(\frac{x_{\rm in}^2}{2\sigma_x}
+\frac{p_{\rm in}^2}{2\sigma_p}\right)\right],
\end{eqnarray}
where $\sigma_x$ and $\sigma_p$ are the variances of the
Q function of the teleported mode for the corresponding quadratures.
Using Eqs.~(\ref{tel}), the fidelity becomes for $g=1$
\begin{eqnarray}\label{fid2}
F=[1&+&(g_{\rm swap}-1)^2(e^{+2r_1}+e^{+2s_1})/4
\nonumber\\
&+&(g_{\rm swap}+1)^2(e^{-2r_2}+e^{-2s_2})/4
\nonumber\\
&+&g_{\rm swap}^2(\eta_c^{-2}-1)
+\eta_a^{-2}-1]^{-1/2}
\nonumber\\
\times[1
&+&(g_{\rm swap}-1)^2(e^{+2r_2}+e^{+2s_2})/4
\nonumber\\
&+&(g_{\rm swap}+1)^2(e^{-2r_1}+e^{-2s_1})/4
\nonumber\\
&+&g_{\rm swap}^2(\eta_c^{-2}-1)
+\eta_a^{-2}-1]^{-1/2} \;.
\end{eqnarray}

For unknown coherent input states, an (average) fidelity 
$F>\case{1}{2}$ is only 
achievable using entanglement \cite{Fuchs}. Thus, if for some $g_{\rm swap}$
(for some local operation on mode 4 by Bob based on 
Claire's results) $F>\case{1}{2}$, entanglement swapping 
must have taken place. 
Otherwise Alice and Bob, who initially did not share entanglement, 
were not able to beat the classical fidelity limit using mode 1 and 4. 
The assumption $g=1$ is the optimal choice for Bob's local operation
based on Alice's results.

Let us first consider four equally squeezed states $r_1=r_2=s_1=s_2=r$.
In this case with unit efficiency ($\eta_c=\eta_a=1$), the fidelity 
is optimized for $g_{\rm swap}=\tanh 2r$ ($g=1$) and becomes
$F_{\rm opt}=(1+1/\cosh 2r)^{-1}$. For any $r>0$, we obtain
$F_{\rm opt}>\case{1}{2}$. For $\eta_c\neq 1$ and $\eta_a\neq 1$, the
optimum gain is $g_{\rm swap}=\sinh 2r/(\cosh 2r + \eta_c^{-2} - 1)$.
For the more general case $r_1=r_2=r$ and $s_1=s_2=s$, 
we find the optimum gain
\begin{eqnarray}\label{optgain}
g_{\rm swap}=\frac{\sinh 2r+\sinh 2s}{\cosh 2r+\cosh 2s+2 
\eta_c^{-2}-2} \;.
\end{eqnarray}
Using this gain we obtain the optimum fidelity
with unit efficiency
\begin{eqnarray}\label{optfid}
F_{\rm opt}=\left\{1+\frac{\cosh[2(r-s)]+1}{\cosh 2r + \cosh 2s}
\right\}^{-1} \;.
\end{eqnarray}
This fidelity is equal to $\case{1}{2}$ and never exceeds the classical
limit if either $r=0$ or $s=0$.
The reduced states of mode 1 and 4 after the detection
of mode 2 and 3 are separable if either $r=0$ or $s=0$ \cite{Peter}.
{\it Both} initial two-mode states need to be squeezed and hence
entangled for entanglement swapping to occur.
Then, for any nonzero squeezing $r>0$ {\it and} $s>0$, we obtain
$F_{\rm opt}>\case{1}{2}$ indicating that entanglement swapping took place.
The reduced states of mode 1 and 4 after detecting
mode 2 and 3 are inseparable for any $r>0$ {\it and} $s>0$ \cite{Peter}.

Let us now consider the case where each of the two initial
entangled states is generated with only one single-mode squeezed state.
We set $r_1=s_1=r$ and $r_2=s_2=0$. With
unit efficiency, we find the optimum gain $g_{\rm swap}=\tanh r$. 
The optimum fidelity then becomes
\begin{eqnarray}\label{optfid2}
F_{\rm opt}&=&\{ [1+2\,e^{+2r}/(e^{+2r}+1)\nonumber\\ 
&+&(\tanh r)^2(\eta_c^{-2}-1)+\eta_a^{-2}-1]
\nonumber\\
&\times&[1+2\,e^{-2r}/(e^{-2r}+1)\nonumber\\ 
&+&(\tanh r)^2(\eta_c^{-2}-1)+\eta_a^{-2}-1]
\}^{-1/2} \;.
\end{eqnarray}
Note that this fidelity can be optimized further for nonunit efficiency, 
as we have only used the optimum gain for unit efficiency.
With unit efficiency ($\eta_c=\eta_a=1$) this fidelity exceeds the classical 
limit $F_{\rm opt}>\case{1}{2}$ for any nonzero squeezing $r_1=s_1=r>0$.
Thus, entanglement swapping can be realized even with only
two single-mode squeezed states, provided that two initial entangled states
are produced. Indeed, the creation of entanglement is possible using
only one single-mode squeezed state for any nonzero squeezing 
\cite{Ah,Matteo,Pvl}.
Therefore we can generally say that any nonzero entanglement in
both of the two initial entanglement sources is sufficient for
entanglement swapping to occur. In order to achieve perfect teleportation
of arbitrary coherent states
with fidelity $F=1$, four infinitely squeezed states $r_1=r_2=s_1=s_2=r
\to\infty$ are necessary and Bob has to perform a displacement with
$g_{\rm swap}=1$ using Claire's results. It is impossible for Alice
and Bob to achieve quantum teleportation of unknown coherent states with
$F>\case{1}{2}$ for $g_{\rm swap}=0$, i.e., without a local operation based 
on the results of Claire's detection. 
The optimum fidelity using mode 1 and 4 after entanglement
swapping is worse than the optimum fidelity using the entanglement
of the initial modes 1 and 2 or 3 and 4. This indicates that the
degree of entanglement after entanglement swapping deteriorates 
compared to the initial entangled states. In figure 2 is shown the 
comparison between the optimum fidelities of coherent-state 
teleportation using entangled states produced from entanglement
swapping and without swapping.

\begin{figure}[htb]
\begin{center}
\begin{psfrags}
     \psfrag{F}[cb]{\Large $F~~~$}
     \psfrag{squeezing}[c]{\Large ~~~~~~~~squeezing [dB]}
     \psfrag{a}[c]{$~a$}
     \psfrag{b}[c]{$b$}
     \psfrag{c}[c]{$~~~~c$}
     \psfrag{d}[c]{$d$}
     \psfrag{e}[c]{$~~e$}
     \psfrag{classical}{\Large \bf classical}
     \psfrag{quantum}{\Large \bf quantum}
\epsfxsize=3.2in
     \epsfbox[0 60 400 320]{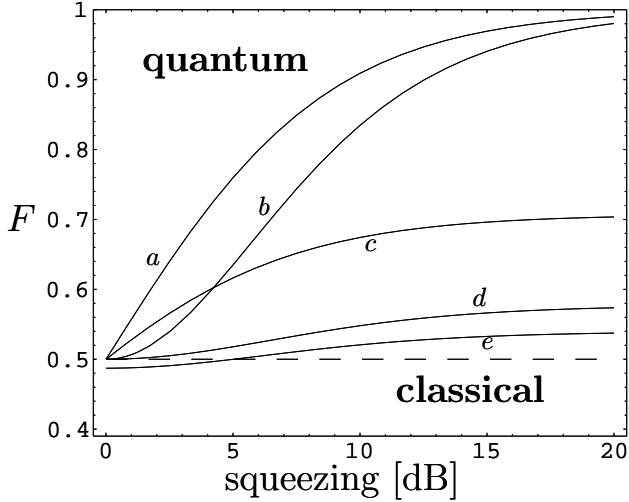}
\end{psfrags}
\end{center}
\caption{Optimum fidelity for the teleportation of an arbitrary coherent state 
         ($g=1$) using a) entanglement produced with two equally
         squeezed single-mode states, b) the output of entanglement 
         swapping with four equally squeezed single-mode states,
         c) entanglement produced with one single-mode squeezed state,
         d) the output of entanglement swapping with two equally squeezed
         single-mode states, e) the state as in d) with
         detector efficiencies $\eta_c^2=\eta_a^2=0.95$.}  
\label{fig2}
\end{figure}

The maximum fidelity achievable using entanglement produced with one 
single-mode squeezed state is $F=1/\sqrt{2}$ \cite{Pvl}.
The maximum fidelity achievable using the output of entanglement 
swapping with two equally squeezed single-mode states is 
$F=1/\sqrt{3}$.
For 6 dB squeezing and detectors with efficiency
$\eta^2=0.99$, the optimum fidelity using the output of entanglement 
swapping with two equally squeezed single-mode states becomes
$F=0.5201$. Squeezing of 10 dB and the same efficiency yields
$F=0.5425$. Here, the gain $g_{\rm swap}=\tanh r$ has been chosen which
is the optimum gain with two equally squeezed single-mode states
for unit efficiency.

Tan proposes continuous-variable entanglement swapping as the teleportation 
of the signal beam of a parametric amplifier using the entanglement 
between signal and idler beam of another parametric amplifier \cite{Tan}.
The entanglement of the teleported signal beam with the idler beam in Tan's
protocol is confirmed by combining them at a beamsplitter and detecting
the $\hat{x}$ quadrature at one output and the $\hat{p}$ quadrature at 
the other output:
$\hat{x}=(\hat{x}_1-\hat{x}_4)/\sqrt{2}, 
\hat{p}=(\hat{p}_1+\hat{p}_4)/\sqrt{2}$.
A violation of the uncertainty relation
$\langle\Delta\hat{x}^2\rangle\langle\Delta\hat{p}^2\rangle\geq 1/16$
proves the entanglement of mode 1 and 4 \cite{Tan}.
A similar, sufficient inseparability criterion for continuous-variable 
systems has been given very recently \cite{Duan} that in the context of our 
entanglement swapping scheme would require the violation of
$\langle(\hat{x}_1-\hat{x}_4)^2\rangle+\langle(\hat{p}_1+\hat{p}_4)^2\rangle
\geq 1$. 

It is obvious that the violation of these inequalities correponds
to a fidelity $F>\case{1}{2}$ in our subsequent coherent-state teleportation.
However, we have directly looked at the separability of the projected
states of mode 1 and 4 after the detection of mode 2 and 3 \cite{Peter}.
Since these states are inseparable for any nonzero entanglement in both
initial sources, they should enable quantum teleportation, as has been shown.
We have demonstrated that in this case, the reduced state $\hat{\rho}_{14}$
can always be transformed by some local displacement
to a state $\hat{\rho}_{14}'$ that provides $F>\case{1}{2}$ in 
coherent-state teleportation and has quadratures violating
$\langle(\hat{x}_1-\hat{x}_4')^2\rangle+\langle(\hat{p}_1+\hat{p}_4')^2
\rangle\geq 1$. We have given the gain $g_{\rm swap}$ of this
displacement for any verification, either by further teleportation
or by simple detection \cite{Polkgain}.

In Tan's scheme \cite{Tan}, unit gain $g_{\rm swap}$ has been chosen.
Therefore entanglement is only confirmed 
if the initial states exceed a certain degree of 
squeezing (e.g. for equally squeezed states 3 dB).
By choosing the right gain entanglement can be verified for any
squeezing, in principle even if the initial entangled states are 
built from the minimal resource of two single-mode squeezed states.
 
Detecting the output and applying Tan's (or Duan et al.'s) 
inequality for verification is easy,
but it requires bringing the entangled subsystems together
and measuring states that are now local. This provides only an indirect
confirmation that the entanglement is preserved through teleportation. 
Our verification leaves
the subsystems separate and directly demonstrates the entanglement by 
exploiting it in a second round of teleportation.

In summary, we have given a protocol for 
continuous-variable entanglement swapping, i.e., for the unconditional 
teleportation of entanglement, using squeezed light and linear optics. 
Entanglement swapping occurs for any nonzero entanglement (any nonzero
squeezing) in both of the two initial entanglement sources.
{\it This can be realized even with only two single-mode squeezed states}.
Our verification scheme would provide a compelling demonstration
of unconditional teleportation of entanglement.

This work was funded by a ``DAAD Doktorandenstipendium im
Rahmen des gemeinsamen HSP III von Bund und
L\"{a}ndern'' and EPSRC Grant No.\ GR/L91344.
PvL thanks Sze M.Tan for sending his manuscript.


\begin{references}
\bibitem{Benn} C.\ H.\ Bennett {\it et al.}, 
Phys.\ Rev.\ Lett.\ {\bf 70}, 1895 (1993).
\bibitem{Vaid} L.\ Vaidmann, Phys.\ Rev.\ A {\bf 49}, 1473 (1994). 
\bibitem{Sam98a} S.\ L.\ Braunstein and H.\ J.\ Kimble, Phys.\ 
Rev.\ Lett.\ {\bf 80}, 869 (1998).
\bibitem{Bou} D.\ Bouwmeester {\it et al.}, Nature {\bf 390}, 575 (1997).
\bibitem{Mart} D.\ Boschi {\it et al.}, Phys.\ Rev.\ Lett.\ {\bf 80}, 1121 
(1998).
\bibitem{Furu} A.\ Furusawa {\it et al.}, Science {\bf 282}, 706 (1998).
\bibitem{Zuk} M.\ Zukowski {\it et al.}, Phys.\ Rev.\ Lett.\ {\bf 71},
4287 (1993).
\bibitem{Pan} J.-W.\ Pan {\it et al.}, Phys.\ Rev.\ Lett.\ {\bf 80},
3891 (1998).
\bibitem{Bose1} S.\ Bose, V.\ Vedral, and P.\ L.\ Knight,
Phys.\ Rev.\ A {\bf 57}, 822 (1998).
\bibitem{Bose2} S.\ Bose, V.\ Vedral, and P.\ L.\ Knight, 
quant-ph/9812013.
\bibitem{Briegel} H.-J.\ Briegel {\it et al.}, Phys.\ Rev.\ Lett.\ {\bf 81},
5932 (1998).
\bibitem{Duer} W.\ D\"{u}r {\it et al.}, Phys.\ Rev.\ A {\bf 59},
169 (1999).
\bibitem{Ralph} R.\ E.\ S.\ Polkinghorne and T.\ C.\ Ralph, 
Phys.\ Rev.\ Lett.\ {\bf 83}, 2095 (1999), quant-ph/9906066.
\bibitem{Tan} S.\ M.\ Tan, Phys.\ Rev.\ A {\bf 60}, 2752 (1999).
\bibitem{Fuchs} S.\ L.\ Braunstein, C.\ A.\ Fuchs, and H.\ J.\ Kimble,
to appear in J.\ Mod.\ Opt., quant-ph/9910030.
\bibitem{Reid} M.\ D.\ Reid and P.\ D.\ Drummond, Phys.\ Rev.\ Lett.\ 
{\bf 60}, 2731 (1988).
\bibitem{Ah} Y.\ Aharonov {\it et al.}, Ann.\ Phys.\ (NY) {\bf 39},
498 (1966).
\bibitem{Matteo} M.\ G.\ A.\ Paris, Phys.\ Rev.\ A {\bf 59}, 1615 (1999).  
\bibitem{Pvl} P.\ van Loock and S.\ L.\ Braunstein,
submitted to Phys.\ Rev.\ Lett., quant-ph/9906021.
\bibitem{Peter} P.\ van Loock and S.\ L.\ Braunstein, in preparation.
\bibitem{Duan} L.-M.\ Duan {\it et al.}, quant-ph/9908056. 
\bibitem{Polkgain} Polkinghorne and Ralph \cite{Ralph} have found a similar 
gain condition which ensures that the correlations of single 
polarization-entangled photons are verified via Bell inequalities
after teleporting them.



\end{references}
\end{document}